\documentclass[conference]{IEEEtran}
\IEEEoverridecommandlockouts
\pdfoutput=1 
\usepackage{cite}
\usepackage{amsmath,amssymb,amsfonts}
\usepackage{algorithmic}
\usepackage{graphicx}
\newtheorem{theorem}{Theorem}
\newtheorem{lemma}{Lemma}
\usepackage{textcomp}
\usepackage{xcolor}
\def\BibTeX{{\rm B\kern-.05em{\sc i\kern-.025em b}\kern-.08em
    T\kern-.1667em\lower.7ex\hbox{E}\kern-.125emX}}
\begin{document}
\bibliographystyle{IEEEtran}
\title{SNR Scaling Laws for Radio Sensing with Extremely Large-Scale MIMO\\}
\author{\IEEEauthorblockN{Huizhi Wang\IEEEauthorrefmark{1} and 
		Yong Zeng\IEEEauthorrefmark{1}\IEEEauthorrefmark{2}}
	
	\IEEEauthorblockA{\IEEEauthorrefmark{1}National Mobile Communications Research Laboratory, Southeast University, Nanjing 210096, China}
	\IEEEauthorblockA{\IEEEauthorrefmark{2}Purple Mountain Laboratories, Nanjing 211111, China}
	\IEEEauthorblockA{Email: wanghuizhi226@163.com, yong\underline{~}zeng@seu.edu.cn.}
}

\maketitle

\begin{abstract}
\textbf{Mobile communication networks were designed to mainly support ubiquitous wireless communications, yet they are expected to also achieve radio sensing capabilities in the near future. Most prior studies on radar sensing focus on distant targets, which usually rely on far-field assumption with uniform plane wave (UPW) models. However, with ever-increasing antenna size, together with the growing need to also sense nearby targets, the far-field assumption may become invalid. This paper studies radar sensing with extremely large-scale (XL) antenna arrays, where a generic model that takes into account both spherical wavefront and amplitude variations across array elements is developed. Furthermore, new closed-form expressions of the sensing signal-to-noise ratios (SNRs) are derived for both XL-MIMO radar and XL-phased-array radar modes. Our results reveal that different from the conventional UPW model where the SNR scales linearly and unboundedly with $N$ for MIMO radar and with $MN$ for phased-array radar, with $M$ and $N$ being the transmit and receive antenna numbers, respectively, more practical SNR scaling laws are obtained. For XL-phased-array radar with optimal power allocation, the SNR increases with $M$ and $N$ with diminishing returns, governed by new parameters called the transmit and receive $\emph{angular spans}$. On the other hand, for XL-MIMO radar, while the same SNR scaling as XL-phased-array radar is obeyed for $N$, the SNR first increases and then decreases with $M$.}
\vspace{-3ex}
\end{abstract}

\section{Introduction}
\vspace{-0.5ex}
With the fifth-generation (5G) mobile communication networks being deployed, researchers have started the ambitious envision of the sixth-generation (6G) networks~\cite{b26,b11,b10}. There is no doubt that 6G should continue to offer wireless communication as its main service, with significant performance improvement in terms of e.g., coverage, connectivity density, communication rate, etc\cite{b26}\cite{b11}. In addition, it is also believed that 6G is quite promising to support ubiquitous localization and radar sensing~\cite{b12,b16,b17}, thanks to the continuous expansion of cellular bandwidth and antenna size that makes high-resolution localization and sensing possible. Besides being offered as new services, 
sensing may also be leveraged to enhance wireless communications~\cite{b18}. Therefore, the integration of sensing and communication has received significant research interest recently, under various terms like joint communication and (radar/radio) sensing (JCAS)~\cite{b13}, joint radar and communication (JRC)~\cite{b14}, and integrated sensing and communication (ISAC)~\cite{b20}.

Multiple-input multiple-output (MIMO) is a key technology for both communication and radar sensing. It is widely known that MIMO communication offers the fundamental spatial multiplexing gain and diversity gain~\cite{b19}. There are similar gains for radar sensing with multiple antennas, corresponding to two different radar modes, namely {\it MIMO radar} and {\it phased-array radar}\cite{b6}. For MIMO radar mode, orthogonal waveforms are transmitted from different antennas, so as to obtain the waveform diversity gain. On the other hand, for phased-array radar, coherent waveforms are transmitted from multiple antennas, so as to obtain high transmit coherent processing gain by beamforming. Compared with phased-array radar, MIMO radar can obtain a larger virtual aperture by applying sparse array, which improves spatial resolution in angle measurement. On the other hand, phased-array radar achieves higher beamforming gain by transmitting and receiving both with narrow beams, but its sensing area during each pulse is limited.

In order to reap the full benefits of multiple antennas, MIMO communications have been tremendously advanced from small MIMO in 4G to massive MIMO in 5G\cite{b21}. Looking forward towards 6G, there have been growing interests in the study of extremely large-scale MIMO (XL-MIMO) communications\cite{b2,b29,b23,b3,b24}, for which the antenna size is so large that conventional assumptions, such as far-field propagation with uniform plane wave (UPW) models, are no longer valid. Instead, the more generic spherical wavefront and variations of signal amplitudes across array elements need to be taken into account. On the other hand, existing studies on multi-antenna radar sensing are still mainly relying on the conventional UPW models\cite{b25}, which is justified by the dominating prior applications on sensing distant targets and/or the use of moderate-size antennas. With the ever-increasing antenna size at base stations (BSs), together with the growing need to also sense nearby targets, it is necessary to develop new sensing models and provide more accurate performance analysis, without restricting to the conventional UPW models. This motivates our current work. 

In this paper, we study radar sensing under the XL-MIMO setup. By discarding the conventional far-field assumption with UPW models, we develop a more generic model that takes into account both spherical wavefront and amplitude variations across array elements. Furthermore, new closed-form expressions are derived for the sensing SNRs for both XL-MIMO radar and XL-phased-array radar modes. Our results reveal that different from the conventional UPW model where the SNR scales linearly and unboundedly with $N$ for MIMO radar and with $MN$ for phased-array radar, with $M$ and $N$ being the transmit and receive antenna numbers, respectively, more practical SNR scaling laws are obtained. For XL-phased-array radar with optimal power allocation, the SNR increases with $M$ and $N$ with diminishing returns, governed by new parameters called the transmit and receive $\emph{angular spans}$\cite{b2}. On the other hand, for XL-MIMO radar, while the same SNR scaling as XL-phased-array radar is obtained for $N$, the SNR first increases and then decreases with $M$. We also show that our developed model and performance analysis include the conventional UPW counterpart as special cases.

\begin{figure}[htbp]
	\vspace{-3.5ex}
\setlength{\abovecaptionskip}{-0.2cm}
\setlength{\belowcaptionskip}{-0.5cm}
	\centerline{\includegraphics[width=3.9in,height=2.3in]{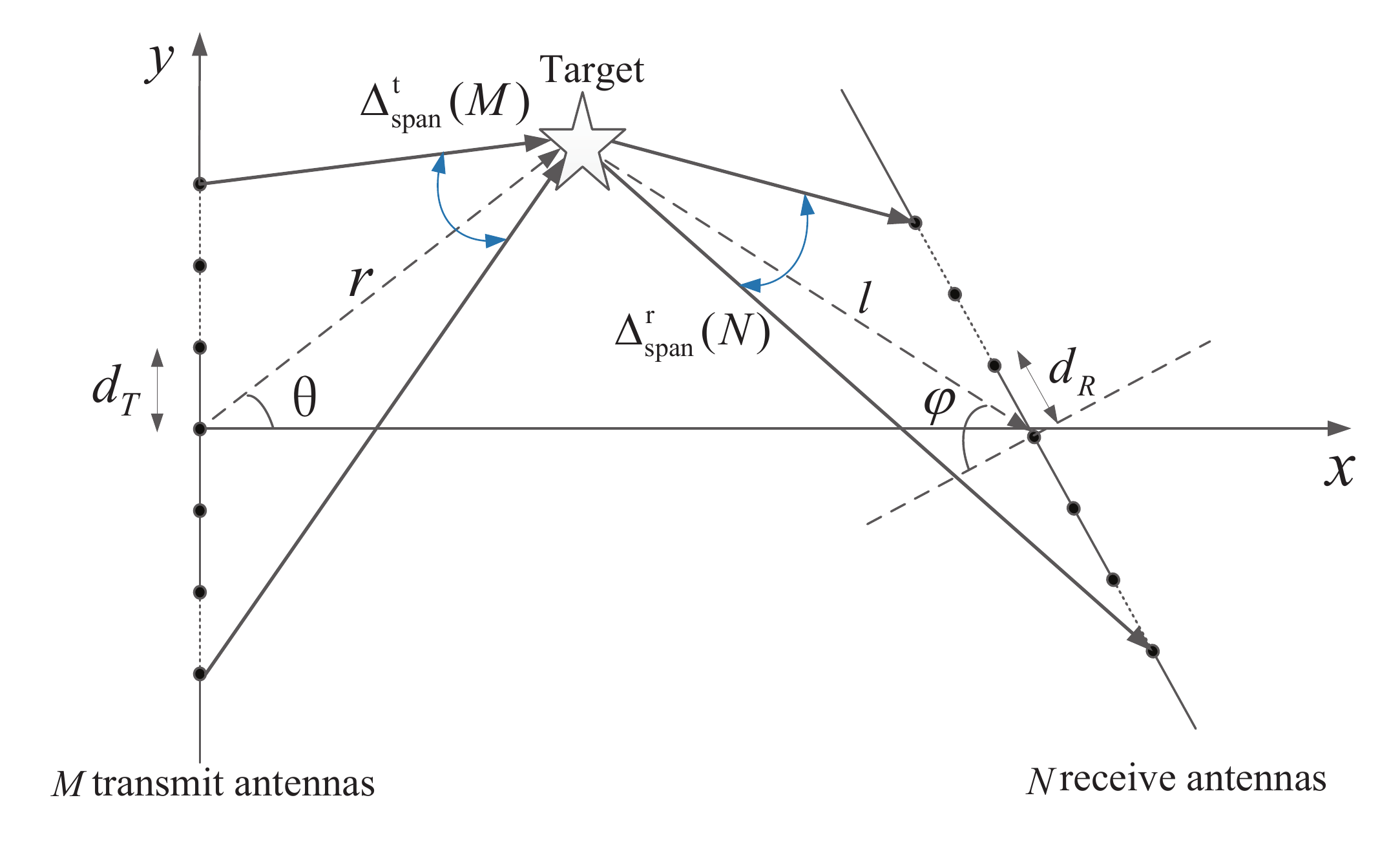}}
	\caption{Radar sensing with extremely large-scale MIMO. }
	\label{fig1}
	\vspace{-2ex} 
\end{figure}
\section{SYSTEM MODEL}
As shown in Fig.\ref{fig1}, we consider a bistatic radar sensing system with extremely large-scale antenna arrays at the transmitter and receiver. Let $M\gg 1$ and $N\gg 1$ denote the number of transmit and receive antenna elements, respectively. For notational convenience, we assume that both $M$ and $N$ are odd numbers. Furthermore, we assume uniform linear arrays (ULAs) at both the transmitter and receiver, and their inter-element spacing is denoted as $d_T$ and $d_R$, respectively. Without loss of generality, we consider a Cartesian coordinate system where the transmit ULA is placed along the $y$-axis and centered at the origin. Therefore, the location of the $m$th transmit element is$\ {{\bf{w}}_m} = {[0,m{d_T}]^T},\ $where$\ m \in \mathcal{M}$, with $\mathcal{ M }  \triangleq  \left\{ {0, \pm 1, \cdots , \pm {{(M - 1)} \mathord{\left/
			{\vphantom {{(M - 1)} 2}} \right.
			\kern-\nulldelimiterspace} 2}} \right\}.\ $Let $\ {\bf{q}} = {[r\cos \theta ,r\sin \theta ]^T}\ $ denote the location of the radar target, where $r$ is the distance between the target and the center of the transmit array and $\ \theta  \in \left[ { - {\pi  \mathord{\left/
	{\vphantom {\pi  {2,{\pi  \mathord{\left/
						{\vphantom {\pi  2}} \right.
						\kern-\nulldelimiterspace} 2}}}} \right.
	\kern-\nulldelimiterspace} {2,{\pi  \mathord{\left/
			{\vphantom {\pi  2}} \right.
			\kern-\nulldelimiterspace} 2}}}} \right]\ $is the direction of the target with respect to the normal vector of the transmit array. Therefore, the distance between the target and the $m$th transmit antenna is 
\begin{equation}
\setlength\abovedisplayskip{3pt}
\setlength\belowdisplayskip{2pt}
\begin{aligned}
	{r_m} = \left\| {{{\bf{w}}_m}{\rm{ - }}{\bf{q}}} \right\| &= \sqrt {r_{}^2 - 2rm{d_T}\sin (\theta ) + {{(m{d_T})}^2}} \\ 
	&= r\sqrt {1 - 2m{\varepsilon _T}\sin (\theta ) + {m^2}{\varepsilon _T}^2} ,
\end{aligned}
\label{eq9}
\end{equation}
where $\ {\varepsilon _T} \buildrel \Delta \over = \frac{{{d_T}}}{r} \ll 1,\ $since $d_T$ is typically on wavelength scale that is much smaller than $r$.

Note that (\ref{eq9}) is a generic distance expression that includes the conventional array modelling based on the far-field assumption $r\gg Md_T$ as special cases. For XL-MIMO system where the far-field assumption no longer holds, the exact distance expression (\ref{eq9}) is necessary to accurately model the signal phase and amplitude relationships across array elements. 
In this case, the transmit array response vector is not just a function of direction $\theta$, but also depends on distance $r$, which can be expressed as\cite{b2}
\begin{equation}
	  \setlength\abovedisplayskip{4pt}
	\setlength\belowdisplayskip{4pt}
\begin{aligned}
\ {\bf{a}}(r,\theta ) = {[{a_{\scriptscriptstyle{- \frac{{M - 1}}{2}}}}(r,\theta ), \cdots ,{a_m}(r,\theta ), \cdots ,{a_{\frac{{M - 1}}{2}}}(r,\theta )]^T},\
\end{aligned}
\label{eq2}
\end{equation}
where $\ {a_m}(r,\theta ) = \frac{{\sqrt {{\beta}} }}{{{r_m}}}{e^{ - j\frac{{2\pi }}{\lambda }{r_m}}},\ $with $\ {{\beta}}\ $ denoting the channel power gain at the reference distance of $r=1$ m.

Similarly, let $l$ denote the distance between the target and the center of the receive antenna array, and $\varphi$ denote the direction of the target with respect to the normal vector of the receive array. Then the receive array response vector can be similarly expressed as
\begin{equation}
	\setlength\abovedisplayskip{4pt}
	\setlength\belowdisplayskip{4pt}
\begin{aligned}
\ {\bf{b}}(l,\varphi ) = {[{b_{ - \frac{{N - 1}}{2}}}(l,\varphi ), \cdots ,{b_n}(l,\varphi ), \cdots ,{b_{\frac{{N - 1}}{2}}}(l,\varphi )]^T},\
\end{aligned}
\label{eq3}
\end{equation}
where$\ \small {b_n}(l,\varphi )=\frac{{\sqrt {{\beta}} }}{{{l_n}}}{e^{ - j\frac{{2\pi }}{\lambda }{l_n}}},\ $with$\ {l_n} = l\sqrt {1 - 2n{\varepsilon _R}\sin (\varphi ) + {n^2}{\varepsilon _R}^2},\ $
where$\ {\varepsilon _R} \buildrel \Delta \over = \frac{{{d_R}}}{l} \ll 1.\ $

Let $\mathbf x(t)\in \mathbb{C}^{M\times 1} $ denote the transmitted waveform. Then the received signal $\mathbf r(t)\in \mathbb{C}^{N\times 1} $ due to the target reflection can be expressed as
\begin{equation}
			  \setlength\abovedisplayskip{2pt}
	\setlength\belowdisplayskip{3pt}
\ {\bf{r}}(t) = \kappa {\bf{b}}(l,\varphi ){{\bf{a}}^T}(r,\theta ){\bf{x}}(t - \tau ) + {\bf{z}}(t),\
\label{eq11}
\end{equation}
where $\kappa$ is a complex reflection coefficient that includes the impact of radar cross section (RCS) of the target, $\tau$ is the propagation delay of the scattered signal by the target, and $\ {\bf{z}}(t)\in \mathbb{C}^{N\times 1}$ is the independent and identically distributed (i.i.d.) additive white Gaussian noise (AWGN) with zero mean and variance$\ {\sigma ^2}.\ $

In the following, we present the performance analysis of the MIMO radar and phased-array radar modes with extremely large-scale antenna arrays, which we term as {\it XL-MIMO radar} and {\it XL-phased-array radar}. 
\section{Performance Analysis and SNR Scaling Laws}
\vspace{-1ex}
\subsection{XL-MIMO Radar}
For MIMO radar, no beam is formed at the transmitter side. Instead, the entire area of interest is uniformly illuminated during each dwell by transmitting orthogonal waveforms from the $M$ transmit antennas\cite{b6}. In this case, the transmitted waveform $\mathbf x(t)$ in (\ref{eq11}) is
\begin{equation}
		  \setlength\abovedisplayskip{1pt}
\setlength\belowdisplayskip{2pt}
\small
\ {{\bf{x}}}(t) = \sqrt {\frac{P}{M}} {\bf{s}}(t),\
\label{eq18}
\end{equation}
where $P$ is the total transmit power, and $\ {\bf{s}}(t)=[s_m(t)]_{m\in \mathcal M} \ $ represents the $M$ orthogonal waveforms, i.e.,\cite{b9}
\begin{equation}
		  \setlength\abovedisplayskip{4pt}
\setlength\belowdisplayskip{4pt}
\ \int_{{T_p}} {{s_m}(t)s_k^*(t - \alpha )} dt = \left\{ \begin{array}{l}
	{R_{ss}}(\alpha ),m = k,\\
	0,\ \ \ \ \ \ \ m \ne k,
\end{array} \right.\
\end{equation}
where $T_p$ is the radar pulse width, and $\ {R_{ss}}(\alpha)\ $ is the autocorrelation function of the waveforms $s_m(t)$, with ${R_{ss}(0)}=1$. Note that in (\ref{eq18}), an equal power $P/M$ is allocated for each antenna, since in MIMO radar mode, the transmitter has no a priori information about the target and thus no power optimization can be applied. By substituting (\ref{eq18}) into (\ref{eq11}), the received signal for XL-MIMO radar is
\begin{equation}
\setlength\abovedisplayskip{3pt}
\setlength\belowdisplayskip{3pt}
{\small
\begin{aligned}
	{\bf{r}}(t) &= \kappa \sqrt {\frac{P}{M}} {\bf{b}}(l,\varphi ){{\bf{a}}^T}(r,\theta ){\bf{s}}(t - \tau ) + {\bf{z}}(t)\\
	&= \kappa \sqrt {\frac{P}{M}} {\bf{b}}(l,\varphi )\sum\nolimits_{m =  - \frac{{M - 1}}{2}}^{\frac{{M - 1}}{2}} {{a_m}(r,\theta ){s_m}(t - \tau )}  + {\bf{z}}(t).
\end{aligned}}
\hspace{-1.8ex}
\end{equation}
By applying matched filtering to ${\bf{r}}(t)$ with each of the orthogonal waveforms$\ s_k(t - \alpha ),\ $$k\in \mathcal {M}$, where $\alpha$ is some selected time delay that may be different from the groundtruth delay $\tau$, the output signal can be expressed as 
\begin{equation}
\setlength\abovedisplayskip{2pt}
\setlength\belowdisplayskip{1pt}
{\small
\begin{aligned}
{{\bf{y}}_{k}} &= \int\limits_{{T_p}} {{\bf{r}}(t)} s_k^*(t - \alpha )dt\\ 
&= \kappa \sqrt {\frac{P}{M}} {\bf{b}}(l,\varphi )\sum\limits_{m =  - \frac{{M - 1}}{2}}^{\frac{{M - 1}}{2}} {{a_m}(r,\theta )} \int\limits_{{T_p}} {{s_m}(t - \tau )s_k^*(t - \alpha )dt} \\
&+ {\tilde {\bf{z}}_k}\\
&=\kappa \sqrt {\frac{P}{M}} {\bf{b}}(l,\varphi ){a_k}(r,\theta ){R_{ss}}(\alpha  - \tau ) + {\tilde {\bf{z}}_k},
\end{aligned}}
\end{equation}
where $\ {{{\bf{\tilde z}}}_k} \buildrel \Delta \over = \int\limits_{{T_p}} {{\bf{z}}(t)} s_k^*(t - \alpha )dt.\ $By concatenating $\mathbf y_{k}\in \mathbb{C}^{N\times 1}$ for all $k\in \mathcal {M}$, we obtain the following $MN$ dimensional data vector
\begin{equation}
\setlength\abovedisplayskip{0.5pt}
\setlength\belowdisplayskip{3pt}
\ {\bf{y}} = \kappa \sqrt {\frac{P}{M}} {\bf{b}}(l,\varphi ) \otimes {\bf{a}}(r,\theta ){R_{ss}}(\alpha  - \tau ) + {\bf{\tilde z}},\
\label{eq5}
\end{equation}
where $\ {\bf{\tilde z}}=[\tilde{\mathbf z}_k]_{k\in \mathcal M}\in \mathbb{C}^{MN\times 1} $. It can be shown that $\tilde{\mathbf z}$ follows the distribution $\tilde{\mathbf z}\sim \mathcal{CN}(\mathbf 0, \sigma^2 \mathbf I_{MN})$. Based on (\ref{eq5}), receive beamforming can be applied to the $MN$-dimensional data $\mathbf y$ for target searching. Note that different from the conventional UPW MIMO radar where the steering vector of the virtual MIMO is only direction-dependent, that for XL-MIMO depends on both target direction and distances. Therefore, the receive beamforming can be applied to search for both target direction and distance, specified by the parameters $(r',\theta')$ and $(l',\varphi')$. The corresponding receive beamforming vector is $\ \small{{\bf{v}}^H} = \frac{{{{\bf{b}}^H}(l',\varphi ')}}{{\left\| {{\bf{b}}(l',\varphi ')} \right\|}} \otimes \frac{{{{\bf{a}}^H}(r',\theta ')}}{{\left\| {{\bf{a}}(r',\theta ')} \right\|}},\ $which results in
\begin{equation}
\setlength\abovedisplayskip{3pt}
\setlength\belowdisplayskip{3pt}
\small
\begin{aligned}
y &= {{\bf{v}}^H}{\bf{y}} \\
&= \frac{{\kappa {R_{ss}}(\alpha  - \tau )}}{{\left\| {{\bf{b}}(l',\varphi ')} \right\|\left\| {{\bf{a}}(r',\theta ')} \right\|}}\sqrt {\frac{P}{M}} \big( {{{\bf{b}}^H}(l',\varphi ') \otimes {{\bf{a}}^H}(r',\theta ')} \big) \\
&\times {\bf{b}}(l,\varphi ) \otimes {\bf{a}}(r,\theta ) + {{\bf{v}}^H}{\bf{\tilde z}}.
\end{aligned}
\label{eq46}
\end{equation}
When the searching parameters match with that of the target, i.e.,$\ \theta' = \theta,$ $\varphi' = \varphi,$ $l' = l,$ $r' = r, $ and$\ \tau  = \alpha,\ $the output SNR is maximum. In this case, (\ref{eq46}) reduces to
\begin{equation}
		\setlength\abovedisplayskip{4pt}
	\setlength\belowdisplayskip{3pt}
	\small
	\begin{aligned}
\ y = \kappa \sqrt {\frac{P}{M}} \left\| {{\bf{b}}(l,\varphi )} \right\|\left\| {{\bf{a}}(r,\theta )} \right\| + \frac{{\left( {{{\bf{b}}^H}(l,\varphi ) \otimes {{\bf{a}}^H}(r,\theta )} \right)}}{{\left\| {{\bf{b}}(l,\varphi )} \right\|\left\| {{\bf{a}}(r,\theta )} \right\|}}{\bf{\tilde z}}.\
	\end{aligned}
\label{eq47}
\end{equation}
The corresponding peak sensing SNR is then given by
\begin{equation}
		  \setlength\abovedisplayskip{3pt}
\setlength\belowdisplayskip{3pt}
\ {\gamma _{{\rm{XL - MIMO}}}} = \frac{{P{{\left| \kappa  \right|}^2}}}{{{\sigma ^2}M}}{\left\| {{\bf{b}}(l,\varphi )} \right\|^2}{\left\| {{\bf{a}}(r,\theta )} \right\|^2}.\
\label{eq19}
\end{equation}
\begin{theorem}
For XL-MIMO radar with $M$ transmit and $N$ receive antennas, the peak output SNR can be expressed in closed-form as
\begin{equation}
	\setlength\abovedisplayskip{2pt}
	\setlength\belowdisplayskip{2pt}
	\begin{aligned}
\ {\gamma _{{\rm{XL - MIMO}}}} = \frac{{P{{\left| \kappa  \right|}^2}{\beta ^2}\Delta _{{\rm{span}}}^{\rm{t}}(M)\Delta _{{\rm{span}}}^{\rm{r}}(N)}}{{{\sigma ^2}M{d_T}{d_R}rl\cos (\theta )\cos (\varphi )}},\
	\end{aligned}
\label{gamma_mimo}
\end{equation}
where $\ \Delta _{{\rm{span}}}^{\rm{t}}(M) = \arctan \left( {{{M{d_T}} \mathord{\left/
			{\vphantom {{M{d_T}} {{\rm{2}}r\cos \theta }}} \right.
			\kern-\nulldelimiterspace} {{\rm{2}}r\cos \theta }} - \tan \theta } \right) + \arctan \left( {{{M{d_T}} \mathord{\left/
			{\vphantom {{M{d_T}} {{\rm{2}}r\cos \theta }}} \right.
			\kern-\nulldelimiterspace} {{\rm{2}}r\cos \theta }} + \tan \theta } \right)\ $and$\ \Delta _{{\rm{span}}}^{\rm{r}}(N) = \arctan \left( {\frac{{N{d_R}}}{{{\rm{2}}l\cos \varphi }} - \tan \varphi } \right) + \arctan \left( {\frac{{N{d_R}}}{{{\rm{2}}l\cos \varphi }} + \tan \varphi } \right)\ $are termed as the transmit and receive {\it angular spans}\cite{b2}, respectively, corresponding to the angles formed by the line segments connecting the target with both ends of the transmit/receive arrays, as illustrated in Fig.\ref{fig1}. 
\end{theorem}

\begin{IEEEproof}
	The proof of Theorem 1 follows from the same technique as Appendix A of \cite{b2}, where the closed-form expressions for$\ {\left\| {{\bf{a}}(r,\theta )} \right\|^2}\ $and$\ {\left\| {{\bf{b}}(l,\varphi )} \right\|^2}\ $in (\ref{eq19}) are obtained by integration, using the facts that $\varepsilon_T\ll 1$ and $\varepsilon_R\ll 1$. 
\end{IEEEproof}

Theorem 1 shows that for any fixed number of transmit antnenas $M$, the SNR increases with the number of receive antennas $N$, but with diminishing return that is governed by the receive angular span $\ {\Delta _{{\rm{span}}}^{\rm{r}}(N)}.\ $ On the other hand,  when $N$ is fixed, the SNR first increases and then decreases with $M$. This is expected since for MIMO radar mode, the total transmit power $P$ needs to be equally divided between the $M$ transmit antennas. When $M$ is small, as $M$ increases, the SNR increases since the increase of$\ {\Delta _{{\rm{span}}}^{\rm{t}}(M)}\ $dominates the power division factor $M$, while the reverse is true as $M$ exceeds certain number. Consider the extreme case when $M\rightarrow \infty$, we have$\ \Delta _{{\rm{span}}}^{\rm{t}}(M) \to \pi \ $and$\ \mathop {\lim }\limits_{M \to \infty } {\gamma _{{\rm{XL - MIMO}}}} = 0.\ $ 

For a direct comparison, the peak SNR for conventional UPW model can be expressed as\cite{b27}
\begin{equation}
\setlength\abovedisplayskip{3pt}
\setlength\belowdisplayskip{3pt}
\ {\gamma _{{\rm{UPW - MIMO}}}} = \frac{{P{{\left|\kappa \right|}^2}{\beta ^2}N}}{{{\sigma ^2}{r^2}{l^2}}},\
\label{eq6}
\end{equation}
which is independent of $M$, and scales linearly and unboundedly with $N$. The difference between (\ref{gamma_mimo}) and (\ref{eq6}) can be explained by the more accurate modelling of the signal amplitude across array elements in the former model. Specifically, for UPW model, equal amplitude is assumed across array elements. This leads to linear increase of the receive beamforming gain. By contrast, with the more general XL-MIMO model, the signal amplitudes at different elements depend on their actual distances with the target. Therefore, as the number of antennas $N$  increases, the additional antenna will have further distance with the target, and thus results in diminishing beamforming gains as in (\ref{gamma_mimo}).  

In the following, we consider some special cases of Theorem 1 to gain further insights. Lemma \ref{lem1} to Lemma \ref{lem3} below can be proved by following similar techniques as \cite{b2}. The details are omitted due to space limitations. 
\begin{lemma}\label{lem1}
	When$\ r \gg \frac{{M{d_T}}}{2},\ $the SNR in (\ref{gamma_mimo}) reduces to
	\begin{equation}
				  \setlength\abovedisplayskip{3pt}
		\setlength\belowdisplayskip{3pt}
\ {\gamma _{{\rm{XL - MIMO}}}} = \frac{{P{{\left| \kappa  \right|}^2}{\beta ^2}\Delta _{{\rm{span}}}^{\rm{r}}(N)}}{{{\sigma ^2}{d_R}{r^2}l\cos (\varphi )}}.\
	\end{equation}
\end{lemma}

\begin{lemma}{\label{lem2}}
When$\ l \gg \frac{{N{d_R}}}{2},\ $the SNR in (\ref{gamma_mimo}) reduces to
\begin{equation}
			  \setlength\abovedisplayskip{3pt}
	\setlength\belowdisplayskip{3pt}
\ {\gamma _{{\rm{XL - MIMO}}}} = \frac{{P{{\left| \kappa  \right|}^2}N{\beta ^2}\Delta _{{\rm{span}}}^{\rm{t}}(M)}}{{{\sigma ^2}{d_T}M{l^2}r\cos (\theta )}}.\
\end{equation}
\end{lemma}

\begin{lemma}{\label{lem3}}
When$\ {r} \gg \frac{{M{d_T}}}{2}\ $and$ \ {l} \gg \frac{{N{d_R}}}{2},\ $the SNR in (\ref{gamma_mimo}) reduces to 
\begin{equation}
\setlength\abovedisplayskip{-1pt}
\setlength\belowdisplayskip{3pt}
\ {\gamma_{\rm{XL-MIMO}}} \approx {\gamma_{\rm{UPW-MIMO}}} = \frac{{P{{\left| \kappa  \right|} ^2}{\beta ^2}N}}{{{\sigma ^2}{r^2}{l^2}}}.\
\end{equation}
\end{lemma}

Lemma \ref{lem3} shows that when the distances between the target and transmit/receive arrays are large enough, the SNR of our generic model reduces to the conventional UPW model. 
\vspace{-1.5ex}
\subsection{XL-Phased-Array Radar}
\vspace{-1ex}
For phased-array radar, transmit beamforming is also formed towards certain direction $\theta'$ and distance $r'$. In this case, the transmitted signal in (\ref{eq11}) can be expressed as
\begin{equation}
\setlength\abovedisplayskip{3pt}
\setlength\belowdisplayskip{3pt}
\ {\bf{x}}(t) = \sqrt {\mathrm{diag}\{ {\bf{c}}\} } {{\bf{a}}^*}(r',\theta ')s(t),\
\label{eq12}
\end{equation}
where $\mathbf c\in \mathbb{R}^{M\times 1}$ is a power allocation coefficient vector across the $M$ antennas,$\ {{\bf{a}}^*}(r',{\theta'})\ $is the transmit steering vector towards the target at distance $r'$ and direction $\theta',$ $s(t)$ is the single transmitted waveform satisfying$\ \int_{{T_p}} {s(t)} {s^*}(t-\alpha)dt = R(\alpha),\ $where $R(\alpha)$ is the autocorrelation function.
Note that with (\ref{eq12}), the transmit power of antenna $m$ is
\begin{equation}
			  \setlength\abovedisplayskip{3pt}
	\setlength\belowdisplayskip{3pt}
\ {P_m} = {C_m}\frac{\beta }{{r_m^{'2}}},\
\label{cm}
\end{equation}
where $C_m$ is the element in $\mathbf c$. With the total power  $P$, $\ C_m\ $should satisfy:$\ \sum \nolimits_{m =  - \frac{{M - 1}}{2}}^{\frac{{M - 1}}{2}} {\frac{{{C_m}}}{{r_m^{'2}}}}  = \frac{P}{\beta }.\ $By substituting (\ref{eq12}) into (\ref{eq11}), the received signal for XL-phased-array radars is
\begin{equation}
		  \setlength\abovedisplayskip{5pt}
\setlength\belowdisplayskip{5pt}
\ {\bf{r}}(t) = \kappa {\bf{b}}(l,\varphi ){{\bf{a}}^T}(r,\theta )\sqrt {\mathrm{diag}\{ {\bf{c}}\} } {{\bf{a}}^*}(r',\theta ')s(t-\tau) + {\bf{z}}(t).\	
\end{equation}
By applying matched filtering for$\ {\bf{r}}(t)\ $with the transmitted waveform$\ s(t - \alpha ),\ $where $ \alpha$ is some selected delay, we obtain
\begin{equation}
\setlength\abovedisplayskip{2pt}
\setlength\belowdisplayskip{2pt}
\small
\begin{aligned}
{\bf{y}}&=\int_{{T_p}} {{\bf{r}}(t){s^*}(t - \alpha )} dt \\
&= \kappa {\bf{b}}(l,\varphi ){{\bf{a}}^T}(r,\theta )\sqrt {\mathrm{diag}\{ {\bf{c}}\} } {{\bf{a}}^*}(r',\theta ')R(\alpha  - \tau ) + {\bf{\tilde z}},
\hspace{-1ex}
\end{aligned}
\end{equation}
with $\ {\bf{\tilde z}} \buildrel \Delta \over = \int\limits_{{T_P}} {{\bf{z}}(t-\alpha)} {s^*}(t)dt,\ $which can be shown to follow the distribution $\tilde{\mathbf z}\sim \mathcal{CN}(\mathbf 0, \sigma^2 \mathbf I_{N})$. After receive beamforming with beamforming vector$\ {{\bf{v}}^H} = \frac{{{{\bf{b}}^H}(l',\varphi ')}}{{\left\| {{\bf{b}}(l',\varphi ')} \right\|}},\ $we have
\begin{equation}
			  \setlength\abovedisplayskip{1pt}
	\setlength\belowdisplayskip{1pt}
	\small
	\begin{aligned}
y &=\mathbf v^H \mathbf y\\
&= \frac{{\kappa {{\bf{b}}^H}(l',\varphi '){\bf{b}}(l,\varphi )}}{{\left\| {{\bf{b}}(l',\varphi ')} \right\|}}{{\bf{a}}^T}(r,\theta )\sqrt {\mathrm{diag}\{ {\bf{c}}\} } {{\bf{a}}^*}(r',\theta ')R(\alpha  - \tau ) \\
&+\frac{{{{\bf{b}}^H}(l',\varphi ')}}{{\left\| {{\bf{b}}(l',\varphi ')} \right\|}}{\bf{\tilde z}}.
	\end{aligned}
\label{eq22}
\end{equation}
When the parameters match with the groundtruth values, i.e., $\theta' = \theta,$ $\varphi' = \varphi,$ $l' = l,$ $r' = r,$ $\tau  = \alpha ,\ $we obtain
\begin{equation}
			  \setlength\abovedisplayskip{2pt}
	\setlength\belowdisplayskip{2pt}
	\begin{aligned}
\ y = \kappa \left\| {{\bf{b}}(l,\varphi )} \right\|\beta \sum\limits_{m =  - \frac{{M - 1}}{2}}^{\frac{{M - 1}}{2}} {\frac{{\sqrt {{C_m}} }}{{r_m^2}}}  + \frac{{{{\bf{b}}^H}(l,\varphi )}}{{\left\| {{\bf{b}}(l,\varphi )} \right\|}}{\bf{\tilde z}}.\
	\end{aligned}
\label{eq10}
\end{equation}
The resulting SNR for (\ref{eq10}) is
\begin{equation}
			  \setlength\abovedisplayskip{4pt}
	\setlength\belowdisplayskip{4pt}
	\begin{aligned}
\ {\gamma_{\rm{XL-PH}}} = \frac{{{{\left| \kappa  \right|} ^2}{\beta ^2}{{\left\| {{\bf{b}}(l,\varphi )} \right\|}^{\rm{2}}}}}{{{\sigma ^2}}}{\Big( {\sum\limits_{m =  - \frac{{M - 1}}{2}}^{\frac{{M - 1}}{2}} {\frac{{\sqrt {{C_m}} }}{{r_m^2}}} } \Big)^2}.\
	\end{aligned}
\label{eq23}
\end{equation}
In the following, we derive the SNR for (\ref{eq23}) with equal power allocation and optimal power allocation, respectively.
\subsubsection{Equal power allocation}
With equal power allocation, all the $M$ transmit antennas are allocated with identical power, which is $P/M$. It follows from (\ref{cm}) that$\ {C_m} = \frac{{Pr_m^2}}{{M{\beta}}},\ $$\forall m$. Note that equal power allocation is optimal for conventional UPW modellng since all antennas have equal channel gains. Though suboptimal in general for XL-phased-array radar, equal power allocation is also considered here since it is easy for practical implementation. 
\begin{theorem}{\label{t2}}
For XL-phased-array radar with equal power allocation using $M$ transmit and $N$ receive antennas, the peak output SNR can be expressed as
\begin{equation}
			  \setlength\abovedisplayskip{4pt}
	\setlength\belowdisplayskip{4pt}
\ {\gamma_{\rm{XL-PH}}} = \frac{{P{{\left| \kappa  \right|}^2}{\beta ^2}\Delta _{{\rm{span}}}^{\rm{r}}(N)}}{{M{\sigma ^2}{d_R}l\cos (\varphi )}}{\psi ^2}(M),\
\label{eq20}
\end{equation}
where$\ \psi (M) = \frac{1}{{r{\varepsilon _T}}}\ln \Big( {\frac{{{p_2} + \sqrt {1 + p_2^2} }}{{{p_1} + \sqrt {1 + p_1^2} }}} \Big)\ $, with$\ {p_1} = \frac{{ - \frac{{M{\varepsilon _T}}}{2} - \sin (\theta )}}{{\cos (\theta )}}\ $and$\ {p_2} = \frac{{\frac{{M{\varepsilon _T}}}{2} - \sin (\theta )}}{{\cos (\theta )}}.\ $
\begin{IEEEproof}
Please refer to Appendix A.
\end{IEEEproof}
\end{theorem}
Similar to XL-MIMO radar in Theorem 1, Theorem \ref{t2} shows that for any given $M$, the SNR increases with $N$ governed by the receive angular span $\Delta _{{\rm{span}}}^{\rm{r}}(N)$. On the other hand, with $N$ fixed, the SNR first increases and then decreases with $M$, governed by the new parameter $\psi(M)$.
\subsubsection{Optimal power allocation}
With optimal power allocation, the resulting beamforming coefficient for each transmit element matches with the array response of the desired beam location, so that the output SNR is maximum. In this case, the elements of the power allocation vector is$\ {C_m} = \frac{P}{{{{\left\| {{\bf{a}}(r,\theta )} \right\|}^2}}},\ $$\forall m$. It follows from (\ref{cm}) that the transmit power of each antenna is$ \ {P_m} = \frac{{P\beta }}{{{{\left\| {{\bf{a}}(r,\theta )} \right\|}^2}r_m^2}},\ $i.e., less power is allocated to the antennas that are far from the target.
\begin{theorem}{\label{t3}}
For XL-phased-array radar with optimal power allocation using $M$ transmit and $N$ receive antennas, the peak output SNR can be expressed as
\begin{equation}
			  \setlength\abovedisplayskip{2pt}
	\setlength\belowdisplayskip{2pt}
\ {{\tilde \gamma }_{{\rm{XL - PH}}}} = \frac{{P{{\left| \kappa  \right|}^2}{\beta ^2}\Delta _{{\rm{span}}}^{\rm{t}}\left( M \right)\Delta _{{\rm{span}}}^{\rm{r}}\left( N \right)}}{{{\sigma ^2}{d_T}{d_R}rl\cos (\theta )\cos (\varphi )}}.\
\label{eq13}
\end{equation}
\begin{IEEEproof}
By substituting $\ {C_m} = \frac{P}{{{{\left\| {{\bf{a}}(r,\theta )} \right\|}^2}}}\ $ into (\ref{eq23}), we have
\begin{equation}
	\setlength\abovedisplayskip{2pt}
	\setlength\belowdisplayskip{4pt}
	\small
	\begin{aligned}
		{{\tilde \gamma }_{\mathrm{XL - PH}}} &= \frac{{{{\left| \kappa  \right|}^2}{\beta ^2}{{\left\| {{\bf{b}}(l,\varphi )} \right\|}^{\rm{2}}}}}{{{\sigma ^2}}}{\Big( {\sum\limits_{m =  - \frac{{M - 1}}{2}}^{\frac{{M - 1}}{2}} {\frac{1}{{r_m^2}}\sqrt {\frac{P}{{{{\left\| {{\bf{a}}(r,\theta )} \right\|}^2}}}} } } \Big)^2} \\
		&= \frac{{P{{\left| \kappa  \right|}^2}{{\left\| {{\bf{b}}(l,\varphi )} \right\|}^{\rm{2}}}{{\left\| {{\bf{a}}(r,\theta )} \right\|}^2}}}{{{\sigma ^2}}} \\
		&= \frac{{P{{\left| \kappa  \right|}^2}{\beta ^2}\Delta _{{\rm{span}}}^{\rm{t}}(M)\Delta _{{\rm{span}}}^{\rm{r}}(N)}}{{{\sigma ^2}{d_T}{d_R}rl\cos (\theta )\cos (\varphi )}},
	\end{aligned}
\end{equation}
where the last equality follows from Appendix A of \cite{b2}.
\end{IEEEproof}
\end{theorem}

By comparing (\ref{eq13}) with the SNR of XL-MIMO radar in (\ref{gamma_mimo}), we have$\ {{\tilde \gamma }_{{\rm{XL - PH}}}} = M{\gamma _{{\rm{XL - MIMO}}}},\ $i.e., similar to the conventional UPW model, the SNR of XL-phased-array radar is $M$ times of that for XL-MIMO radar.

In the following, we consider the extreme cases of Theorem 3, the proof of which can be obtained following similar technique as \cite{b2} and is thus omitted due to space limitation. 
\begin{lemma}{\label{lem5}}
When$\ M,N \to \infty \ $, the SNR in (\ref{eq13}) reduces to
\begin{equation}
			  \setlength\abovedisplayskip{2pt}
	\setlength\belowdisplayskip{2pt}
	\begin{aligned}
\ \mathop {\lim }\limits_{M,N \to \infty } {{\tilde \gamma }_{{\rm{SW - PH}}}} = \frac{{P{{\left| \kappa  \right|}^2}{\beta ^2}{\pi ^2}}}{{{\sigma ^2}{d_T}{d_R}rl\cos (\theta )\cos (\varphi )}}.\
	\end{aligned}
\label{eq28}
\end{equation}
\end{lemma}
\begin{lemma}{\label{lem6}}
When$\ {r} \gg \frac{{M{d_T}}}{2}\ $and$\ {l} \gg \frac{{N{d_R}}}{2},\ $the SNR in (\ref{eq13}) reduces to
\begin{equation}
			  \setlength\abovedisplayskip{2pt}
	\setlength\belowdisplayskip{2pt}
\ {{\tilde \gamma }_{{\rm{XL - PH}}}} \approx {{\tilde \gamma }_{{\rm{UPW - PH}}}} = \frac{{P{{\left| \kappa  \right|}^2}{\beta ^2}MN}}{{{\sigma ^2}{r^2}{l^2}}}.\
\label{eq15}
\end{equation}
\end{lemma}

\section{SIMULATION RESULTS}
Numerical results are provided in this section to compare our developed models and the conventional UPW models. Unless otherwise stated, we set $M=N$, $l=r=50$ m,  $d_T=d_R=0.0628$m, and$\ \frac{{P{{\left| \kappa  \right|}^2}{\beta ^2}}}{{{\sigma ^2}}} = 50\ $dB.

\begin{figure}[htbp]
	\vspace{-2ex}
	  \setlength{\abovecaptionskip}{-0.1cm}
	\setlength{\belowcaptionskip}{-0.3cm}
	\centerline{\includegraphics[width=0.48\textwidth]{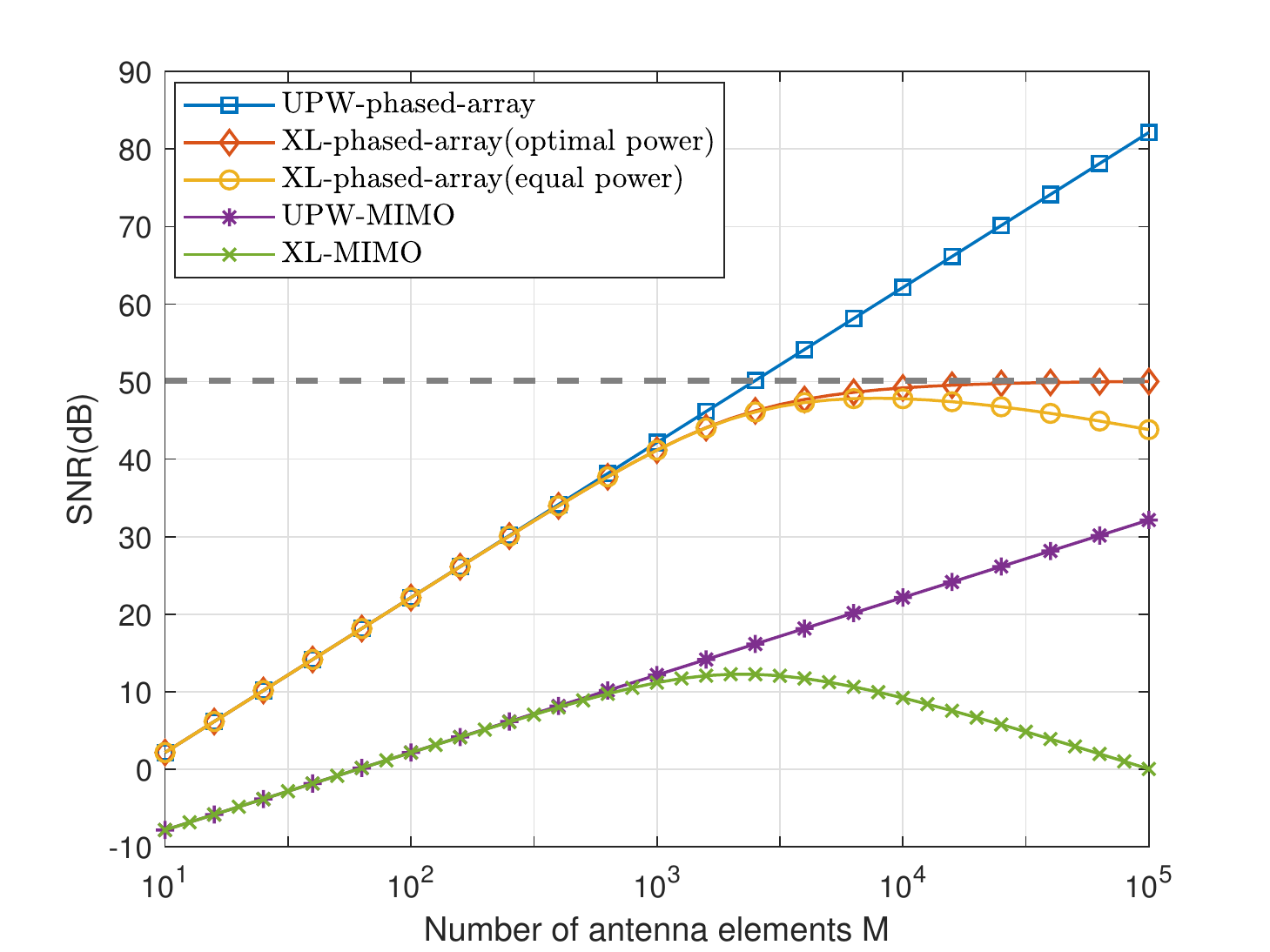}}
	\caption{SNR versus the number of transmit and receive antennas $M=N$ with the conventional UPW model and proposed model.}
	\vspace{-0.5ex} 
	\label{fig2}
\end{figure}

Fig.~\ref{fig2} shows the SNR versus the number of antennas $M$ for the conventional UPW model and the proposed model for both MIMO and phased-array radars. The target direction is set as $\theta=\varphi=0$. It is observed from the figure that for relatively small $M$ values, our proposed model is consistent with the conventional UPW model, which is in accordance with Lemma \ref{lem3} and Lemma \ref{lem6}. However, for moderately large $M$ values, with the conventional far-field UPW model and $M=N$, the SNR increases linearly and unboundedly with $M$ for MIMO radar, due to the receive beamforming gain, and with $M^2$ for phased-array radar, due to both receive and transmit beamforming gains. By contrast, with the proposed model, the SNR increases first and then decreases for XL-MIMO radar and XL-phased-array radar with equal power allocation. On the other hand, for XL-phased-array radar with optimal power allocation, the SNR increases with $M$ and will eventually approach a constant values specified by Lemma \ref{lem5}. Such drastically different SNR scaling laws show the importance of appropriate modelling for radio sensing with XL-MIMO arrays. 
\begin{figure}[htbp]
	  \setlength{\abovecaptionskip}{-0.1cm}
	\setlength{\belowcaptionskip}{-0.3cm}
	\centerline{\includegraphics[width=0.48\textwidth]{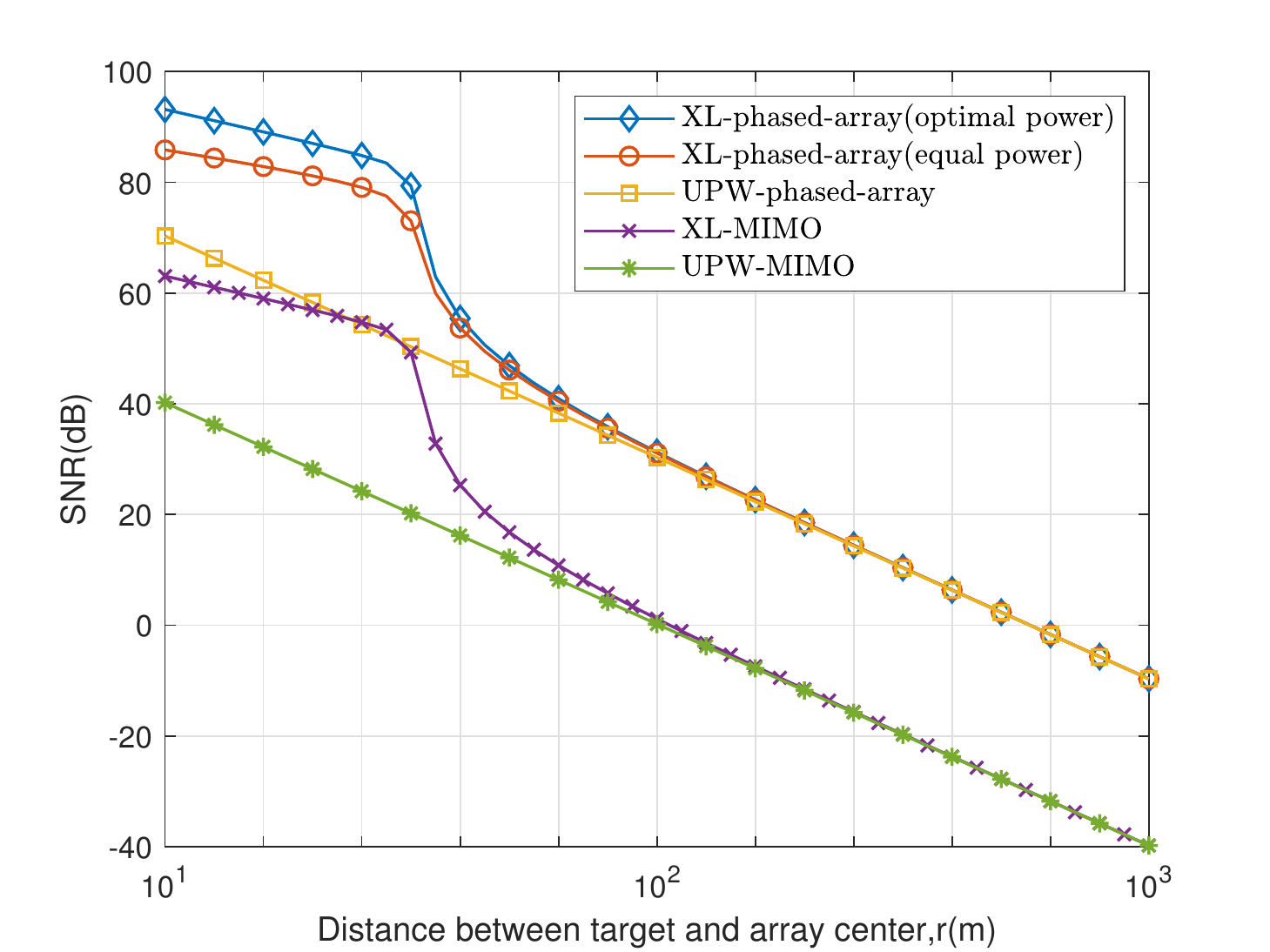}}
	\caption{SNR versus target distance $r$ with the conventional UPW model and proposed model.}
	\vspace{-3.5ex} 
	\label{fig3}
\end{figure}

Fig.~\ref{fig3} plots the SNR versus the distance $r=l$ for different schemes. The number of transmit and receive antenna elements is $M=N=1024$, and the target direction is set to be $\theta=\varphi=88^ \circ $. It is observed that for large target distance $r$, the SNR of the proposed model perfectly matches with the conventional UPW models, but they deviate significantly for relatively small $r$, where far-field assumption no longer holds. In particular, it is observed that for the considered target direction at $88^ \circ$, the conventional UPW model under-estimates the true SNRs, while the reverse is true for $\theta=\varphi=0$, as can be inferred from Fig.~\ref{fig2}. 
\begin{figure}[htbp]
	  \setlength{\abovecaptionskip}{-0.1cm}
	\setlength{\belowcaptionskip}{-0.3cm}
	\centerline{\includegraphics[width=0.48\textwidth]{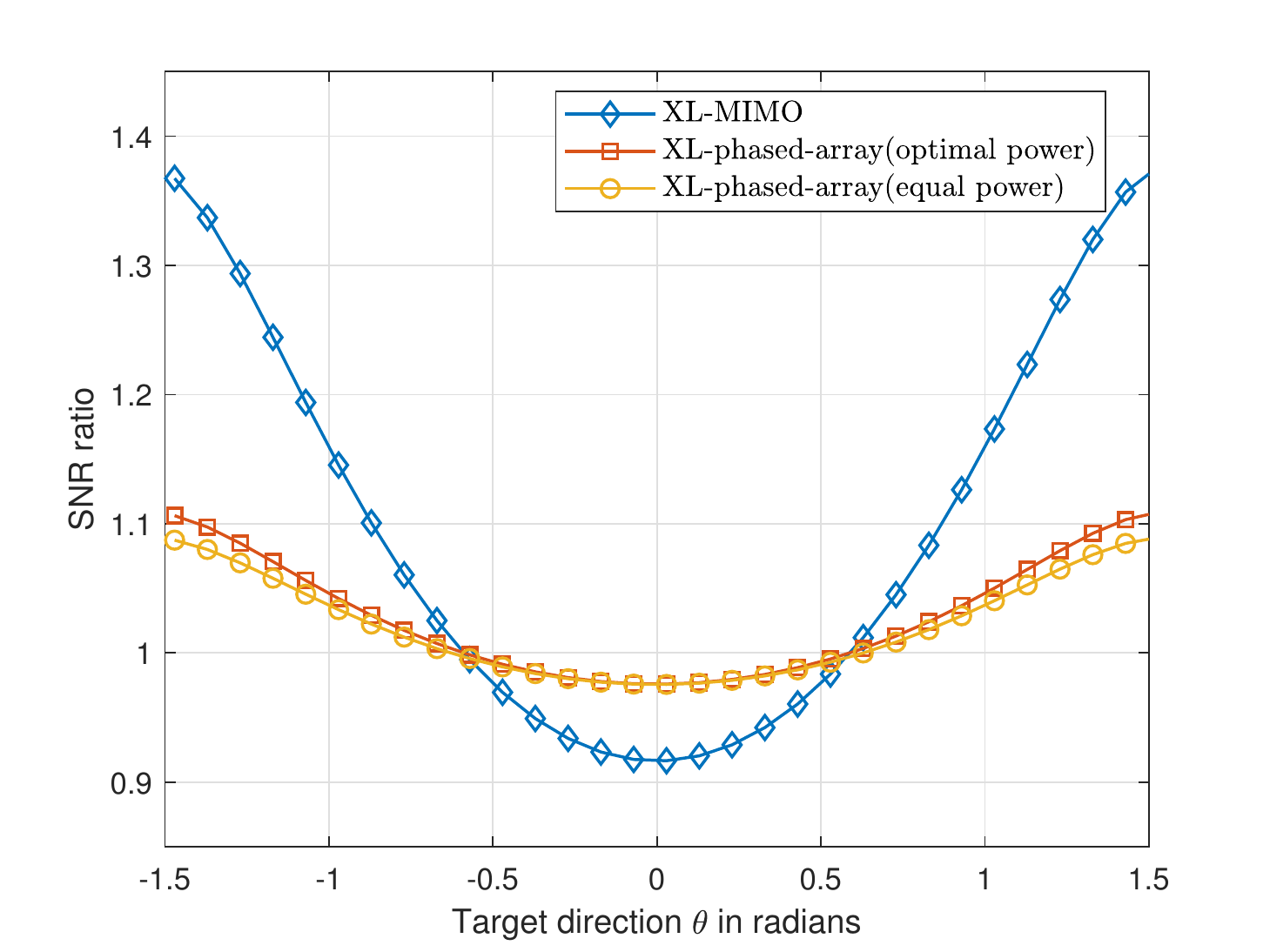}}
	\caption{SNR ratio versus target direction $\theta=\varphi$.}
	\vspace{-3.5ex} 
    \label{fig4}
\end{figure}

To further study the impact of different models, we investigate the SNR ratio\cite{b2}, which is defined as the ratio between the resulting SNR of our proposed model and the conventional UPW model for MIMO-radar and phased-array radar, respectively.
Fig.~\ref{fig4} shows the SNR ratios for different radar modes versus the target direction $\theta=\varphi$. It is observed that for both MIMO radar and phased-array radar, the differences caused by the two models are most significant when the target is at inclined directions, i.e., when $|\theta|$ is large. Furthermore, compared to phased-array operation, MIMO radar will lead to more severe SNR prediction errors if the conventional UPW model is inappropriately used when sensing nearby target and/or when the antenna arrays are large. 
\section{CONCLUSION}
In this paper, we studied the channel modeling and performance analysis for radio sensing with XL-MIMO. By taking into account the spherical wavefront and signal amplitude variations across array elements, new closed-form expressions are derived for XL-MIMO radar and XL-phased-array radar modes. Our results show that the modeling and SNR results based on conventional UPW model are no longer valid in the XL-MIMO scenario. Extensive numerical results are provided to demonstrate the importance of appropriate modelling for sensing with XL-MIMO.
\section*{ACKNOWLEDGMENT}
This work was supported by the National Key R\&D Program of China with grant number 2019YFB1803400.
\section*{APPENDIX A}
By substituting $\ {C_m} = \frac{{Pr_m^2}}{{M{\beta}}}\ $into (\ref{eq23}), we have
\begin{equation}
\setlength\abovedisplayskip{2pt}
\setlength\belowdisplayskip{2pt}
\small
\begin{aligned}
{\gamma _{\rm{XL - PH}}} &= \frac{{{{\left| \kappa  \right|}^2}{\beta ^2}{{\left\| {{\bf{b}}(l,\varphi )} \right\|}^{\rm{2}}}}}{{{\sigma ^2}}}{\Big( {\sum_{m =  - \frac{{M - 1}}{2}}^{\frac{{M - 1}}{2}} {\sqrt {\frac{P}{{r_m^2M\beta }}} } } \Big)^2}\\
&= \frac{{P{{\left| \kappa  \right|}^2}\beta {{\left\| {{\bf{b}}(l,\varphi )} \right\|}^{\rm{2}}}}}{{M{\sigma ^2}}}{\Big( {\sum_{m =  - \frac{{M - 1}}{2}}^{\frac{{M - 1}}{2}} {\frac{1}{{r_m^{}}}} } \Big)^2}\\
&=\frac{{P{{\left| \kappa  \right|}^2}{\beta ^2}\Delta _{{\rm{span}}}^{\rm{r}}(N)}}{{M{\sigma ^2}{d_R}l\cos (\varphi )}}\\
&\times {\Big( {\sum_{m =  - \frac{{M - 1}}{2}}^{\frac{{M - 1}}{2}} {\frac{1}{{r\sqrt {1 - 2m{\varepsilon _T}\sin (\theta ) + {m^2}{\varepsilon _T}^2} }}} } \Big)^2}.
\end{aligned}
	\label{apen1}
\end{equation}
Note that following from Appendix A of \cite{b2}, we have used the identiy$\ {\left\| {{\bf{b}}(l,\varphi )} \right\|^2} = \frac{\beta }{{{d_R}l\cos (\varphi )}}\Delta _{{\rm{span}}}^{\rm{r}}(N).\ $To evaluate the summation inside the parentheses of (\ref{apen1}), we define the function$\ f(x) \buildrel \Delta \over = \frac{1}{{r\sqrt {1 - 2\sin (\theta )x + {x^2}} }},\ $where $\ x \in \left[ { - \frac{{M{\varepsilon _T}}}{2},\frac{{M{\varepsilon _T}}}{2}} \right].\ $Since $\ {\varepsilon _T} \ll 1,\ $we have $\ f(x) \approx f(m{\varepsilon _T}),\ $ $\ \forall x \in \left[ {\left( {m - \frac{1}{2}} \right){\varepsilon _T},\left( {m + \frac{1}{2}} \right){\varepsilon _T}} \right].\ $So we have
\begin{equation}
\setlength\abovedisplayskip{2pt}	\setlength\belowdisplayskip{2pt}
\small
\ \sum_{m =  - \frac{{M - 1}}{2}}^{\frac{{M - 1}}{2}} {f(m{\varepsilon _T}){\varepsilon _T}}  = \int_{ - \frac{{M{\varepsilon _T}}}{2}}^{\frac{{M{\varepsilon _T}}}{2}} {f(x)dx}.\
\label{eq4}
\end{equation}
By substituting $f(x)$ into (\ref{eq4}), the summation in (\ref{apen1}) can be expressed as
\begin{equation}
	\setlength\abovedisplayskip{3pt}
	\setlength\belowdisplayskip{3pt}
	\hspace{-1ex}
	\small
	\begin{aligned}
	\sum\limits_{m =  - \frac{{M - 1}}{2}}^{\frac{{M - 1}}{2}} {\frac{1}{{r_m^{}}}}  &\approx \frac{1}{{r{\varepsilon _T}}}\int_{ - \frac{{M{\varepsilon _T}}}{2}}^{\frac{{M{\varepsilon _T}}}{2}} {\frac{1}{{\sqrt {{x^2} - 2\sin (\theta )x + 1} }}} dt\\
	&\mathop = \limits^{(a)} \frac{1}{{r{\varepsilon _T}}}\int\limits_{{p_1}}^{{p_2}} {\frac{1}{{\sqrt {{p^2} + 1} }}} dp\mathop  = \limits^{(b)} \frac{1}{{r{\varepsilon _T}}}\int\limits_{{\alpha _1}}^{{\alpha _2}} {\frac{1}{{\cos (\alpha )}}} d\alpha \\
	&= \frac{1}{{r{\varepsilon _T}}}\int\limits_{{\alpha _1}}^{{\alpha _2}} {\frac{1}{{1 - {{\sin }^2}(\alpha )}}} d\sin (\alpha )\\
	&= \frac{1}{{2r{\varepsilon _T}}}\left[ {\ln \left| {\frac{{1 + \sin ({\alpha _2})}}{{1 - \sin ({\alpha _2})}}} \right| - \ln \left| {\frac{{1 + \sin ({\alpha _1})}}{{1 - \sin ({\alpha _1})}}} \right|} \right]\\
	&= \frac{1}{{r{\varepsilon _T}}}\ln \left| {\frac{{\tan ({\alpha _2}) + \sqrt {1 + {{\tan }^2}({\alpha _2})} }}{{\tan ({\alpha _1}) + \sqrt {1 + {{\tan }^2}({\alpha _1})} }}} \right|\\
	&= \frac{1}{{r{\varepsilon _T}}}\ln \left| {\frac{{{p_2} + \sqrt {1 + p_2^2} }}{{{p_1} + \sqrt {1 + p_1^2} }}} \right|,
	\end{aligned}
\end{equation}
where $(a)$ follows from the change of variable $\ p = \frac{{x - \sin (\theta )}}{{\cos (\theta )}}\ $with$\ {p_1} = \frac{{ - \frac{{M{\varepsilon _T}}}{2} - \sin (\theta )}}{{\cos (\theta )}}\ $and$\ {p_2} = \frac{{\frac{{M{\varepsilon _T}}}{2} - \sin (\theta )}}{{\cos (\theta )}},\ $and $(b)$ is true by letting$\ p = \tan (\alpha ),\ $with$\ {\alpha _1} = \arctan \Big( {\frac{{ - \frac{{M{\varepsilon _T}}}{2} - \sin (\theta )}}{{\cos (\theta )}}} \Big)\ $and$\ {\alpha _2} = \arctan \Big( {\frac{{\frac{{M{\varepsilon _T}}}{2} - \sin (\theta )}}{{\cos (\theta )}}} \Big).\ $By substituting the above result to (\ref{apen1}), the proof of Thoerem \ref{t2} completes. 

\bibliography{IEEEabrv,ref}
\end{document}